\documentclass[twocolumn]{aastex631}

\usepackage{amsmath}
\usepackage{lipsum}
\usepackage{booktabs}
\usepackage{tabularx}
\usepackage{graphicx}
\usepackage[noabbrev]{cleveref}
\usepackage{ulem}

\submitjournal{ApJ}

\begin{document}

\title{Effects of Bursty Star Formation on [\ion{C}{2}] Line Intensity Mapping of High-redshift Galaxies}

\email{lliu@caltech.edu}

\author[0009-0004-1270-2373]{Lun-Jun Liu}
\affiliation{California Institute of Technology, 1200 E California Blvd, Pasadena, CA 91125, USA}

\author[0000-0003-4070-497X]{Guochao Sun}
\affiliation{CIERA and Department of Physics and Astronomy, Northwestern University, 1800 Sherman Ave, Evanston, IL 60201, USA}

\author[0000-0001-5929-4187]{Tzu-Ching Chang}
\affiliation{NASA Jet Propulsion Laboratory, 4800 Oak Grove Dr, Pasadena, CA 91011, USA}
\affiliation{California Institute of Technology, 1200 E California Blvd, Pasadena, CA 91125, USA}

\author[0000-0002-0658-1243]{Steven~R. Furlanetto}
\affiliation{Department of Physics and Astronomy, University of California, Los Angeles, CA 90095, USA}

\author[0000-0001-5261-7094]{Charles~M. Bradford}
\affiliation{NASA Jet Propulsion Laboratory, 4800 Oak Grove Dr, Pasadena, CA 91011, USA}
\affiliation{California Institute of Technology, 1200 E California Blvd, Pasadena, CA 91125, USA}

\begin{abstract}

Bursty star formation --- a key prediction for high-redshift galaxies from cosmological simulations explicitly resolving stellar feedback in the interstellar medium --- has recently been observed to prevail among galaxies at redshift $z \gtrsim 6$. Line intensity mapping (LIM) of the 158\,$\mu$m [\ion{C}{2}] line as a star formation rate (SFR) indicator offers unique opportunities to tomographically constrain cosmic star formation at high redshift, in a way complementary to observations of individually detected galaxies. To understand effects of bursty star formation on [\ion{C}{2}] LIM, which remain unexplored in previous studies, we present an analytic modeling framework for high-$z$ galaxy formation and [\ion{C}{2}] LIM signals that accounts for bursty star formation histories induced by delayed supernova feedback. We use it to explore and characterize how bursty star formation can impact and thus complicate the interpretation of the [\ion{C}{2}] luminosity function and power spectrum. Our simple analytic model indicates that bursty star formation mainly affects low-mass galaxies by boosting their average SFR and [\ion{C}{2}] luminosity, and in the [\ion{C}{2}] power spectrum it can create a substantial excess in the large-scale clustering term. This distortion results in a power spectrum shape which cannot be explained by invoking a mass-independent logarithmic scatter. We conclude that burstiness must be accounted for when modeling and analyzing [\ion{C}{2}] datasets from the early universe, and that in the extreme, the signature of burstiness may be detectable with first-generation experiments such as TIME, CONCERTO, and CCAT-DSS.

\end{abstract}

\section{Introduction} \label{sec:intro}

The epoch of reionization (EoR) is a frontier for studying the formation and evolution of galaxies. Detailed information about the galaxy populations during the EoR not only sheds light on the exact roles played by galaxies in ionizing the universe, but also provides critical tests for galaxy formation theory. The substantially different cosmological environments in the younger, denser and less metal-enriched high-$z$ universe have enabled various stress tests on the generality and flexibility of models of galaxy formation.    

One of the most intriguing features of high-$z$ ($z \gtrsim 6$) galaxies is the prevalence of bursty star formation, namely the existence of strong time variations in a galaxy's star formation history (SFH), which has been predicted by both simulations and analytic arguments \citep{Dominguez_2015, Sparre_2017, CAFG_2018, FM_2022, Dekel_2023, Li_2023, PF_2023, Sun_2023_OBS} and recently supported by observations made by the James Webb Space Telescope (JWST) either directly \cite[e.g.,][]{Ciesla_2023, Dressler_2023, Looser_2023} or indirectly \cite[e.g.,][]{Shen_2023, Munoz_2023, Sun_2023_LF}. Bursty star formation can have numerous consequences on galaxy formation, affecting a wide range of observables including galaxy spectrum, morphology, chemical abundance, luminosity distribution, and spatial clustering. Nevertheless, despite its clear astrophysical significance, the burstiness of star formation remains challenging to be constrained at the individual galaxy level since we observe galaxies only in snapshots. Assessing the time variability of the SFH from photometric or spectroscopic data is often susceptible to bias from the choice of the SFH prior and model degeneracies. 

Alternatively, to probe the effects of bursty star formation and characterize its significance for galaxy formation at high redshift, one can examine the collective properties and statistics for ensembles of galaxies. As recently demonstrated in \citet{Munoz_2023} for the spatial clustering of galaxies and in \citet{Sun_2023_EBL} for star formation rate (SFR) indicators extractable from the cross-correlation between the extragalactic background light (EBL) and galaxies, effects of bursty star formation strongly impact cosmological signals like galaxy distributions and cosmic background radiations. Therefore, bursty star formation is a key physical process to consider in the modeling and interpretation of high-$z$ cosmological signals. 

Motivated by these studies, we investigate in this work the impact of bursty star formation on another important cosmological probe --- line intensity mapping (LIM), an approach that aims to map the large-scale structure in three dimensions using a spectral line signal without resolving individual sources of emission \cite[see][for a recent review]{BK_2022}. By mapping the aggregate emission of a variety of possible target lines (e.g., \ion{H}{1} 21\,cm, Ly$\alpha$, H$\alpha$, [\ion{C}{2}], CO), LIM has been proposed to be a powerful way to study the astrophysics and cosmology of the high-$z$ universe \citep{Parsons_2022, MasRibas_2023, Sun_2023}. Among the many target lines, the 157.7\,$\mu$m [\ion{C}{2}] line has attracted much attention as a promising probe of cosmic star formation at high redshift, given its strength in typical star-forming galaxies \citep{Hollenbach_1999}, correlation with the SFR \citep{DeLooze_2014, Schaerer_2020} and the CO-dark molecular gas \citep{zanella_2018, vizgan_2022}, and complementarity to the 21\,cm signal during the EoR \citep{Gong_2012, Silva_2015, Yue_2015, Dumitru_2019, YueFerrara_2019, Chung_2020}. However, a large fraction of theoretical studies on [\ion{C}{2}] LIM so far have chosen to employ simple scaling relations that relate [\ion{C}{2}] emission to halo/galaxy properties. While several recent LIM studies have attempted to more systematically address the issue of scatter in [\ion{C}{2}] luminosity based on insights from semi-analytic models \citep{Bethermin_2022} or radiation-hydrodynamic simulations \citep{Murmu_2023}, models that physically connect bursty star formation and LIM signals of SFR tracers are still lacking. Specifically, little effort has been made to quantify the effects of bursty star formation on the [\ion{C}{2}] luminosity distribution and the corresponding LIM signals.

In this paper, we develop an analytic framework for modeling [\ion{C}{2}] luminosity (among other key galaxy properties) in high-$z$ galaxies with bursty SFHs arising from the time delay in supernova feedback. We use this framework to investigate the imprints of bursty star formation and the transition to time-steady star formation on [\ion{C}{2}] LIM signals in a physically motivated manner. For the first time, our analysis proves the importance of properly accounting for the mass-dependent stochasticity in [\ion{C}{2}] luminosity induced by bursty star formation when studying the summary statistics of [\ion{C}{2}] LIM signals at high redshift. Throughout, we assume cosmological parameters consistent with \citet{Planck_2016}.

\section{Models} \label{sec:mod}

Here, we present a simple, analytic framework that links bursty star formation driven by the delayed injection of supernova feedback to [\ion{C}{2}] LIM signals from high-$z$ galaxies. In Section~\ref{sec:sf}, we summarize the essential physics for a self-regulated disk model as the basis for describing galaxy-scale star formation. In Section~\ref{sec:bursty}, we model the evolution of stellar, gas, and metal contents of galaxies, sustained by smooth gas accretion and regulated by supernova feedback injected with a time delay due to stellar evolution. We then use the derived galaxy properties to model the [\ion{C}{2}] emission in Section~\ref{sec:cii}. Next, in Section~\ref{sec:lcii_distr}, we elucidate the connection between time variations of the SFR and [\ion{C}{2}] luminosity for individual halos and approximate the stochasticity of [\ion{C}{2}] luminosity with a log-normal distribution. In Section~\ref{sec:stats}, we apply this functional form of [\ion{C}{2}] luminosity distribution to derive the [\ion{C}{2}] LIM signals. Fiducial values of several key model parameters are provided in Table~\ref{tb:model_params}. 

We emphasize that these models only provide a highly simplified picture of bursty star formation in high-$z$ galaxies that can be incomplete in detail in various ways. We nevertheless consider them to be adequate for physically motivating our phenomenological approach to investigate how bursty star formation affects high-$z$ [\ion{C}{2}] LIM. Some caveats will be pointed out when relevant.

\subsection{The Star-forming Disk--Halo Connection} \label{sec:sf}

To model bursty star formation, we start from a simple description of galaxy-scale stellar and gas dynamics by considering a disk model similar to the ones described in \cite{CAFG_2013} and \cite{CAFG_2018}. For a galaxy hosted by an isothermal halo of virial radius $r_\mathrm{vir}$, velocity dispersion $\sigma$, and circular velocity $v_c = \sqrt{2}\sigma$, we assume that the galactic disk of radius $r_\mathrm{d} = 0.02 r_\mathrm{vir}$ maintains vertical hydrostatic equilibrium through the turbulence driven by stellar feedback and self-regulates to satisfy the Toomre's stability criterion \citep{Toomre_1964}
\begin{equation}
Q = \frac{2 \sigma c_\mathrm{T}}{\pi G \Sigma_\mathrm{g,crit} r_\mathrm{d}} \sim 1, 
\end{equation}
where the turbulence velocity dispersion $c_\mathrm{T}$ is related to the disk orbital frequency $\Omega = v_c / 2\pi r_\mathrm{d}$ and the scale height $h = \Sigma_\mathrm{g}/2 \rho_\mathrm{g}$ by $c_\mathrm{T} \sim \Omega h = \sigma Q f_\mathrm{g} / 2$ \citep{CAFG_2013}. For a Toomre parameter $Q \sim 1$, we can define the critical surface density $\Sigma_\mathrm{g,crit}$ for star formation to occur when gas clouds collapse under gravitational instability (over a free-fall timescale $\tau_\mathrm{ff}$) to be
\begin{equation}
\Sigma_\mathrm{g,crit} \sim  \frac{3 \sigma f_\mathrm{g} \Omega}{2 G}. 
\label{eq:SigmaGas}
\end{equation}
For our calculations, a gas fraction $f_\mathrm{g} = M_\mathrm{g}/M_\mathrm{h} \sim 0.1$ is assumed, consistent with the results from cosmological zoom-in simulations of galaxies at cosmic dawn (Sun et al. in prep.). Star formation occurs only when the disk surface density exceeds the threshold, namely $\Sigma_\mathrm{g} = 0.5 M_\mathrm{g} / \pi r^2_\mathrm{d} > \Sigma_\mathrm{g,crit}$, where $M_{\mathrm{g}}$ is the gas mass. The free-fall timescale, $\tau_\mathrm{ff}$, of the disk is related to the orbital timescale $t_\mathrm{orb} = \Omega^{-1}$ by
\begin{equation}
\tau_\mathrm{ff} = \left( \frac{3 Q}{64\sqrt{2}} \right)^{1/2} t_\mathrm{orb} \approx 0.2 t_\mathrm{orb} 
\end{equation}
for $Q \sim 1$. We note that while JWST has enabled detailed studies of galaxy morphologies at cosmic dawn \citep{Ferreira_2023, Kartaltepe_2023}, it remains to be confirmed whether $z \gtrsim 6$ galaxies can be well approximated by the simple disk model considered. We defer a more thorough investigation of this issue to future work.  

\begin{table}[t!]
\caption{Definition and fiducial values of model parameters}
\begin{tabularx}{\columnwidth}{ccc}
\toprule
\toprule
Parameter & Definition & Value (Reference) \\
\hline
$\epsilon_{\mathrm{ff}}$ & cloud-scale SFE & 0.015 (Eq.~\ref{eq:sfr}) \\
$\eta_{0}$ & mass-loading factor & 5 (Eq.~\ref{eq:mass-loading}) \\
$t_{\mathcal{D},\mathrm{min}}$ & minimum SN delay time & 5\,Myr (Eq.~\ref{eq:SNdelay}) \\
$t_{\mathcal{D},\mathrm{max}}$ & maximum SN delay time & 30\,Myr (Eq.~\ref{eq:SNdelay}) \\
$f_{[\mathrm{CII}]}$ & fraction of carbon in [\ion{C}{2}] & 0.5 (Eq.~\ref{eq:LCII}) \\
$T_{\mathrm{g}}$ & PDR gas temperature & 100\,K (Eq.~\ref{eq:LCII}) \\
$n_{\mathrm{g}}$ & PDR gas density & 3000\,$\mathrm{cm^{-3}}$ (Eq.~\ref{eq:LCII}) \\
$y_{Z}$ & metal yield factor & 0.02 (Eq.~\ref{eq:metal_ode}) \\
\bottomrule
\end{tabularx}
\label{tb:model_params}
\end{table}

\subsection{Bursty Star Formation from Delayed Feedback} \label{sec:bursty}

To model high-$z$ bursty star-forming galaxies, we adopt the analytic model introduced by \cite{FM_2022}\footnote{See also \citet{Orr_2019} for a similar investigation of delayed feedback and SFR variations at later cosmic times.}, assuming that the burstiness of star formation in early galaxies is primarily caused by the delayed injection of supernova feedback. Due to the non-negligible delay time in the explosion of supernovae as a result of stellar evolution (which varies across and within types of supernova), the feedback strength varies strongly over time and thus induces SFR fluctuations. As detailed below, feedback-regulated star formation with a time delay can be described by the solution to a system of delay differential equations (DDEs).

Let $M_\mathrm{h}$, $M_\mathrm{g}$, $M_\mathrm{*}$, and $M_\mathrm{w}$ be the halo mass, gas mass, stellar mass, and the mass of wind-driven outflows, we can describe their time evolution by
\begin{equation}
\dot{M}_\mathrm{h}(t) = \dot{M}_\mathrm{a,h}(t)
\label{eq:mar}
\end{equation}
\begin{equation}
\dot{M}_\mathrm{g}(t) = f_\mathrm{b} \dot{M}_\mathrm{a,h}(t) - \dot{M}_{\star}(t) - \dot{M}_\mathrm{w}(t)
\label{eq:mgas}
\end{equation}

\begin{equation}
\dot{M}_\mathrm{\star}(t) =
\begin{cases}
    \frac{\epsilon_\mathrm{ff}}{\tau_\mathrm{ff}(t)} M_\mathrm{g}(t) & \text{if } \Sigma_\mathrm{g} > \Sigma_\mathrm{g,crit} \\
    0 & \text{otherwise}
\end{cases}
\label{eq:sfr}
\end{equation}

\begin{equation}
\dot{M}_\mathrm{w}(t) = \eta(M_\mathrm{h}, t) \dot{M}^\mathcal{D}_\mathrm{\star}(t-t_\mathcal{D}).
\label{eq:SNdelay}
\end{equation}
Equation~(\ref{eq:mar}) describes the continuous growth of dark matter halos via smooth accretion, the rate of which, $\dot{M}_\mathrm{a,h}$, is determined by the abundance matching technique that conserves the number density of halos \citep{Furlanetto_2017, Furlanetto_2021}. Equation~(\ref{eq:mgas}) describes the regulation of the gas reservoir, with accreted baryons ($f_\mathrm{b} M_\mathrm{a,h}$) removed by both star formation\footnote{Strictly speaking, one should account for the gas return fraction ($\sim1/4$) as a result of stellar evolution, which is neglected here given its minor effect.} and galactic winds driven by supernova feedback. Equation~(\ref{eq:sfr}) describes star formation: we assume that stars form out of the gas reservoir at some given cloud-scale star formation efficiency (SFE), $\epsilon_\mathrm{ff}$, 
per free-fall time, $\tau_\mathrm{ff}$, as long as the gas surface density surpasses the critical value given by Equation~(\ref{eq:SigmaGas}). Equation~(\ref{eq:SNdelay}) gives the rate of galactic outflows driven by the momentum injected by supernovae with a mass-loading factor 
\begin{equation}
\eta(M_\mathrm{h}, z) = \eta_{0} \left( \frac{M_\mathrm{h}}{10^{11.5}\,M_{\odot}} \right)^{-1/3} \left( \frac{1+z}{9} \right)^{-1/2},
\label{eq:mass-loading}
\end{equation}
which multiplies with $\dot{M}^\mathcal{D}_\mathrm{\star}$, the \textit{time-averaged effective} SFR corresponding to the supernova feedback delayed over a time interval $t_{\mathcal{D},\mathrm{min}} < t_\mathcal{D} < t_{\mathcal{D},\mathrm{max}}$ due to stellar evolution (Table~\ref{tb:model_params}). The normalization factor $\eta_0$ is determined by matching the implied UV luminosity functions to observations \citep{Furlanetto_2021}, which yields the galaxy-scale SFE consistent with constraints from abundance matching analysis \citep{SF_2016, Mirocha_2017, SA_2023}. 

The galaxy gas-phase metallicity, $Z \equiv M_{Z} / M_{\mathrm{g}}$, required for modeling [\ion{C}{2}] emission (see next subsection) is self-consistently derived from the solution to the DDEs by tracing the evolution of metal mass, $M_{Z}$, in the ISM. Assuming pristine inflows, we follow \citet{Furlanetto_2021} and express the metal mass evolution as
\begin{equation}
\begin{split}
\dot{M}_{Z} & = (1 - f_\mathrm{m}) y_{Z} \dot{M}_{\star} - Z \dot{M}_{\star} - \eta Z \dot{M}_{\star}^{\mathcal{D}} \\
& = y_{Z} \dot{M}_{\star} - (M_Z/M_{\mathrm{g}}) \dot{M}_{\star} - (M_Z/M_{\mathrm{g}}) \eta \dot{M}_{\star}^{\mathcal{D}},
\end{split}
\label{eq:metal_ode}
\end{equation}
where in the source term $y_{Z} = 0.02$ is the metal yield factor\footnote{The exact value of $y_Z$ can vary by an order of unity depending on the stellar metallicity and the initial mass function; see e.g., \cite{Benson_2010}.} and a fraction $f_\mathrm{m}$ of the metals produced are injected directly into feedback-driven winds launched with a time delay. For simplicity, we assume $f_\mathrm{m}$ is negligible and note that its impact on our models is fully degenerate with $y_Z$. We note that Equation~(\ref{eq:metal_ode}) also assumes that pre-supernova winds rather than supernova-driven winds dominate the metal enrichment. This may be plausible if high-$z$, low-mass galaxies cannot retain the high-velocity supernova ejecta but the low-velocity ejecta due to pre-supernova winds launched by massive stars. However, we have verified that if the opposite is true, replacing $y_{Z} \dot{M}_{\star}$ with $y_{Z} \dot{M}_{\star}^{\mathcal{D}}$ in Equation~(\ref{eq:metal_ode}) does not impact our results in any significant way. A more thorough analysis of the metal enrichment process in early bursty galaxies will be presented in a forthcoming paper (Furlanetto et al. in prep.).

In contrast to the bursty scenario, we also consider a non-bursty, quasi-equilibrium model of galaxy formation, where Equation~(\ref{eq:SNdelay}) and (\ref{eq:metal_ode}) simplify to
\begin{equation}
\dot{M}_\mathrm{w}(t) = \eta(M_\mathrm{h}, t) \dot{M}_{\star}(t), 
\label{eq:sfr_eq}
\end{equation}
\begin{equation}
\dot{M}_{Z} = [y_{Z} - (1 + \eta) Z] \dot{M}_{\star}
\label{eq:metal_eq}
\end{equation}
such that the momentum injection from supernovae happens instantaneously without a delay. This describes the limit when either $t_\mathcal{D}$ approaches zero or the bursty-to-steady transition of star formation completes as a galaxy becomes massive enough. 

\subsection{[\ion{C}{2}] Emission} \label{sec:cii}

In the limit that [\ion{C}{2}] is optically thin, we can express the [\ion{C}{2}] luminosity as \citep{Crawford_1985,Munoz_2014,MunozOh2016}

\begin{equation}
    L_\mathrm{[CII]} = \frac{M_\mathrm{g} \mathrm{(C/H)}}{m_\mathrm{p}} \frac{2\, f_\mathrm{[CII]} A_\mathrm{[CII]} k_\mathrm{B} T_\mathrm{[CII]}}{2 + e^{T_\mathrm{[CII]} / T_\mathrm{g}}(1+n^\mathrm{crit}_\mathrm{[CII]}/n_\mathrm{g})},
\label{eq:LCII}
\end{equation}
where C/H $= 1.5 \times 10^{-4} Z$ \citep{Munoz_2014} is the carbon abundance that scales with metallicity, $m_\mathrm{p}$ is the proton mass, $A_\mathrm{[CII]} = 2.3\times10^{-6}\,\mathrm{s}^{-1}$ is the Einstein coefficient of [\ion{C}{2}] emission, $T_\mathrm{[CII]} = h \nu_{[\mathrm{CII}]}/k_\mathrm{B} = 91\,$K is the [\ion{C}{2}] equivalent temperature, $T_{\mathrm{g}} = 100\,$K is the typical gas temperature for photodissociation regions (PDRs), $n^\mathrm{crit}_\mathrm{[CII]} \approx 3000\,\mathrm{cm^{-3}}$ is the critical density for [\ion{C}{2}], and $n_\mathrm{g}$ is the PDR gas density. Note that even if the optical depth $\tau \gg 1$ for [\ion{C}{2}], Equation~(\ref{eq:LCII}) is still a reasonable approximation when $n_\mathrm{g} \lesssim n^\mathrm{crit}_\mathrm{[CII]}$, a plausible condition of the PDR in high-$z$ galaxies \citep[][]{Valtchanov_2011,Vallini_2020}, because all photons produced can ultimately escape and give rise to an effectively optically thin (EOT) scenario when collisional de-excitation does not dominate over spontaneous emission\footnote{Since observations of both high-redshift galaxies and their low-redshift analogs have found PDR densities to primarily span the range of $10^1$--$10^4\,\mathrm{cm^{-3}}$ \cite[e.g.,][]{Wardlow_2017,Cormier_2019}, we consider the optically thin approximation to be sufficient for our purpose, although it is also noteworthy that an increased PDR density provides one possible explanation for the observed ``[\ion{C}{2}] deficit'' in some galaxies \citep{Luhman_2003}.} \citep{Goldsmith_2012}. The fraction of carbon in the singly ionized phase, $f_\mathrm{[CII]}$, depends on detailed photoionization conditions of the ISM including density, temperature, the interstellar radiation field (ISRF), and the cosmic ray background. In the presence of a substantial cosmic ray background that might be common for galaxies at high redshift, $f_\mathrm{[CII]}$ is expected to be well above 10\% for typical PDR conditions assumed here and therefore we adopt $f_\mathrm{[CII]} = 0.5$ throughout, an approximation consistent with the predictions for galaxies at the EoR from \cite{Olsen_2017}. 

Note that our goal here is to come up with a physically motivated approximation for [\ion{C}{2}] emission that broadly matches the existing high-$z$ observations, rather than modeling in detail the (sophisticated) physics of line production for which our simplistic model is not suited. Nevertheless, in addition to the comparison with observations shown in Section~\ref{sec:result}, we discuss caveats and possible ways of extending the current framework to more reliably model [\ion{C}{2}] emission in Section~\ref{sec:discussion}.

\subsection{[\ion{C}{2}] Luminosity Distribution of Bursty Galaxies} \label{sec:lcii_distr}

Solving the system of DDEs as introduced in Section~\ref{sec:bursty} for any given halo, we can obtain the time evolution of the various galaxy properties of interest, with effects of bursty star formation being manifested as fluctuations around the non-bursty, equilibrium solution assuming no time delay. The [\ion{C}{2}] luminosity, $L_{\mathrm{[CII]}}$, which we calculate with Equation~(\ref{eq:LCII}), also exhibits strong fluctuations when star formation is bursty. As will be shown, for individual halos, the variation of $L_{\mathrm{[CII]}}$ traces that of the SFR with a time lag due to its direct dependence on the gas mass and metallicity. When burst cycles of star formation gradually damp out as the galaxy grows more massive, the amplitude of [\ion{C}{2}] luminosity fluctuations also decreases and eventually stabilizes to the equilibrium value.

This asymptotic behavior of $L_{\mathrm{[CII]}}$ during the bursty-to-steady transition of star formation motivates us to describe $L_{\mathrm{[CII]}}$ at a given halo mass and redshift by fitting a probability density distribution to the $L_{\mathrm{[CII]}}$ values derived from galaxy properties given by the solution to the DDEs. Specifically, taking $L_{\mathrm{[CII]}}$ for halos of different mass predicted by both the bursty and the equilibrium solutions, we find that $L_\mathrm{[CII]}$ fluctuations can be well-described by the following probability distribution function (PDF)\footnote{For brevity, we drop the redshift dependence in the equations below, but note that all of them are also functions of redshift.}
\begin{equation}
\mathcal{P}(\log L | M_\mathrm{h}) = \frac{\exp \left[ - \frac{(\log L - \langle \log L \rangle)^2}{2 \sigma^2_{L}(M_\mathrm{h})} \right]}{\sqrt{2\pi} \sigma_{L}(M_\mathrm{h})},
\label{eq:pdf}
\end{equation}
where $\langle \log L \rangle \equiv \langle \log L (M_{\mathrm{h}}) \rangle$ is the mean luminosity set to be equal to the equilibrium solution and $\sigma_{L}$ is the (strongly mass-dependent) logarithmic scatter determined by fitting an envelope to the $L_{\mathrm{[CII]}}$ evolution track in different halo mass regimes. This is an essential step to estimate the effect of bursty star formation on [\ion{C}{2}] LIM statistics because, despite the large time variability of galaxy properties and $L_{\mathrm{[CII]}}$ predicted by our model, at any given halo mass and redshift these quantities are actually deterministic (fully set by the initial/boundary condition) rather than stochastic. This phenomenological approach of fitting a log-normal PDF provides a reasonable approximation to the $L_\mathrm{[CII]}$ distribution among the halo population at any given cosmic time as predicted by our analytic model. By construction, it is also guaranteed to reduce to the equilibrium solution when star formation becomes steady.

\subsection{[\ion{C}{2}] LIM with Stochasticity from Bursty SFHs} \label{sec:stats}

\begin{figure*}[ht!]
\centering
\includegraphics[width=\textwidth]{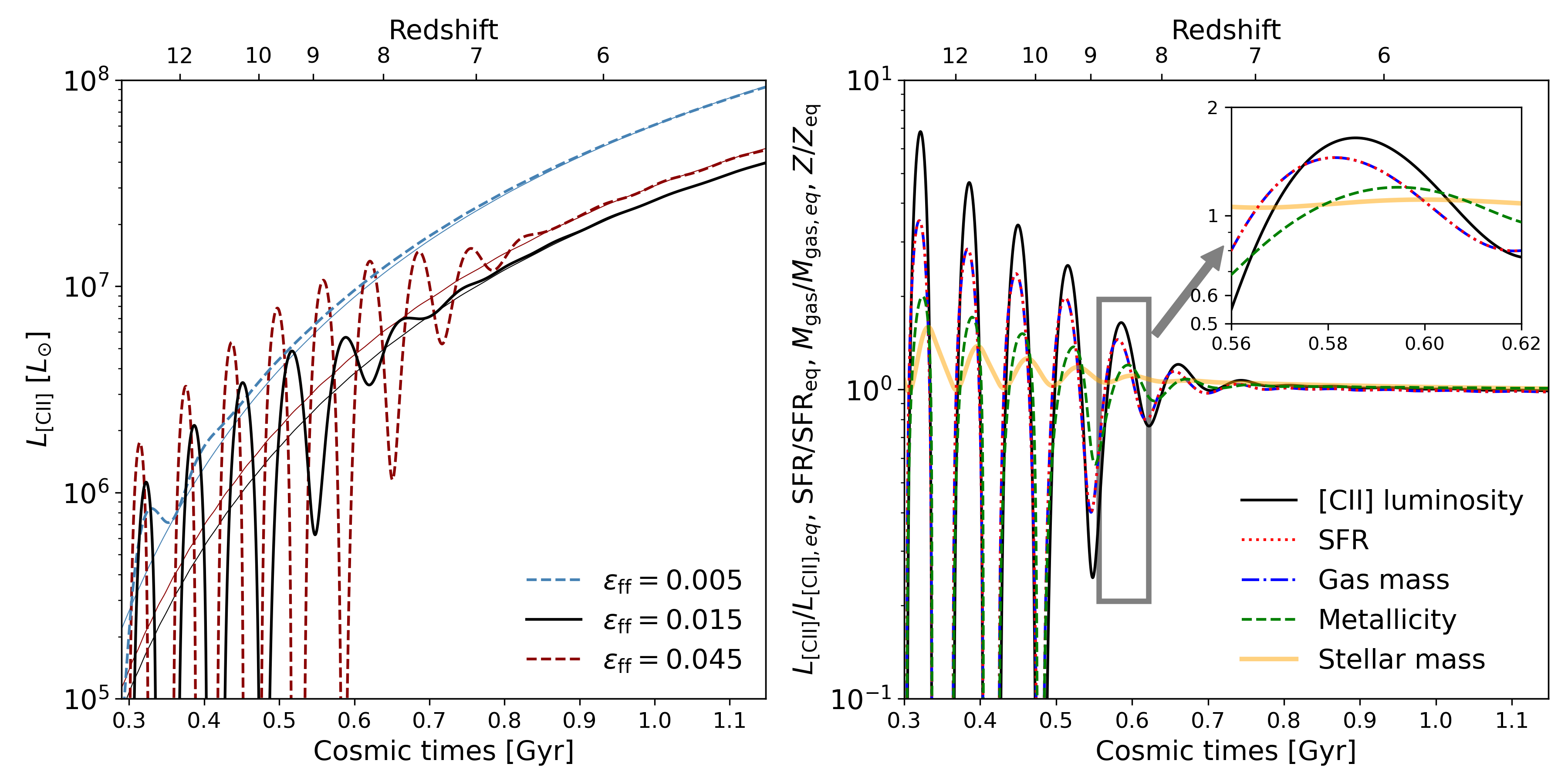}
\caption{Left: comparison of the predicted SFHs, traced by $L_{\mathrm{[CII]}}$, with different levels of burstiness controlled by the SFE $\epsilon_{\mathrm{ff}}$ (0.005, 0.015, and 0.045), for an example halo that grows from $M_{\mathrm{h}} \approx 10^8\,M_{\odot}$ at $z=15$ to $M_{\mathrm{h}} \approx 10^{11}\,M_{\odot}$ at $z=5$. The thick and thin curves represent SFHs predicted by the bursty ($t_\mathcal{D} > 0$) and the equilibrium ($t_\mathcal{D} = 0$) models, respectively, with a transition from bursty to steady star formation visible in the former case. Right: contrast between the bursty SFH and the equilibrium SFH for the same example halo in the fiducial bursty star formation model with $\epsilon_{\mathrm{ff}} = 0.015$. The differences are represented in terms of the ratio of halo properties, including the [\ion{C}{2}] luminosity (black solid), SFR (red dotted), gas mass (blue dash-dotted), metallicity (green dashed), and stellar mass (yellow solid). The inset shows a zoom-in view of these ratios.}
\label{fig:check}
\end{figure*}

Effects of a stochastic $L_\mathrm{[CII]}$--$M_\mathrm{h}$ relation on the summary statistics of [\ion{C}{2}] LIM have been explored in some previous studies \cite[e.g.,][]{Sun_2019,Murmu_2023} and generally included in forecasts of [\ion{C}{2}] LIM signals \cite[e.g.,][]{YueFerrara_2019, Sun_2021, Bethermin_2022, Karoumpis_2022}. However, the physical origin of the stochasticity and its connection to galaxy properties are little explored. Here, for the first time, we employ our bursty star formation models to motivate the parameterization of a stochastic $L_\mathrm{[CII]}$--$M_\mathrm{h}$ relation, whose scatter has a clear physical interpretation pertaining to the variability of $L_\mathrm{[CII]}$ associated with bursty star formation. In what follows, we focus on the [\ion{C}{2}] luminosity function (one-point statistics) and power spectrum (two-point statistics) as the two most important summary statistics for LIM observations and derive their expressions for a stochastic $L_\mathrm{[CII]}$--$M_\mathrm{h}$ relation. 

For each model variation, we use the phenomenological approach explained in Section~\ref{sec:lcii_distr} to describe the distribution of $L_\mathrm{[CII]}$ at any given halo mass and redshift with the log-normal PDF defined by Equation~(\ref{eq:pdf}). Note that in order to account for the intrinsic stochasticity in $L_\mathrm{[CII]}$ (regardless of bursty star formation) due to e.g., variations of the ISM conditions from galaxy to galaxy, we impose a minimum scatter of 0.2\,dex in all cases. A 1$\sigma$ scatter of $\log L_{\mathrm{[CII]}} \gtrsim 0.2$\,dex is consistent with the observed scatter of $L_\mathrm{[CII]}$ at a given SFR \citep{DeLooze_2014, Schaerer_2020}. To find the logarithmic scatter $\sigma_{L}(M_\mathrm{h},z)$ from the best-fit PDF, we compute the halo/galaxy growth history by evolving a wide range of initial halo masses from $z=25$ to $z=5$. This allows us to compile a collection of halo/galaxy evolution tracks and thereby establish a mapping to $\mathcal{P}(\log L | M_\mathrm{h})$ from $M_\mathrm{h}$ and $z$ by fitting a log-normal PDF.

We can then derive the summary statistics by a number of convolutions that involve $\mathcal{P}$. For the observed [\ion{C}{2}] luminosity function, $\Phi = d n / d \log L$, we have
\begin{equation}
\Phi = \int_{M_\mathrm{h,min}} \frac{d n}{d M_\mathrm{h}} \mathcal{P}(\log L | M_\mathrm{h}) d M_\mathrm{h},
\end{equation}
where $d n/d M_\mathrm{h}$ is the halo mass function \citep{Trac_2015} identical to the ones used for deriving the halo accretion rate from abundance matching (Equation~\ref{eq:mar}), and the minimum halo mass for hosting a [\ion{C}{2}]-emitting galaxy is set to $M_\mathrm{h,min} = 10^{9}\,M_{\odot}$ since star formation in less massive halos may be strongly suppressed by reionization feedback \citep{Yue_2016,Dawoodbhoy_2018}. For the [\ion{C}{2}] power spectrum as a function of wavenumber $k$, we break it down into two components, the clustering term and the shot-noise term, namely
\begin{equation}
P(k) = P_\mathrm{clust} + P_\mathrm{shot} = \langle I \rangle^2 \langle b_{L} \rangle^2 P_{\delta \delta}(k) + P_\mathrm{shot},
\end{equation}
where $\langle I \rangle$ is the mean intensity, $\langle b_{L} \rangle$ is the luminosity-weighted bias, and $P_{\delta \delta}(k)$ is the dark matter power spectrum. The individual components are expressed as
\begin{equation}
\langle I \rangle = \int \frac{d n}{d M_\mathrm{h}} \frac{y D^2_A}{4\pi D^2_{L}} d M_\mathrm{h} \int \mathcal{P}(\log L | M_\mathrm{h}) L d \log L,
\end{equation}
\begin{equation}
\langle b_{L} \rangle = \frac{\int \frac{d n}{d M_\mathrm{h}} b(M_\mathrm{h}) d M_\mathrm{h} \int \mathcal{P}(\log L | M_\mathrm{h}) L d \log L}{\int \frac{d n}{d M_\mathrm{h}} d M_\mathrm{h} \int \mathcal{P}(\log L | M_\mathrm{h}) L d \log L},
\end{equation}
and 
\begin{equation}
P_\mathrm{shot} = \int \frac{d n}{d M_\mathrm{h}} \left(\frac{y D^2_A}{4\pi D^2_{L}} \right)^2 d M_\mathrm{h} \int \mathcal{P}(\log L | M_\mathrm{h}) L^2 d \log L,
\label{eq:pshot}
\end{equation}
where $b(M_\mathrm{h})$ is the halo bias, $D_{L}$ and $D_{A}$ are the luminosity and comoving angular diameter distances, respectively, and $y = d \chi / d \nu$ maps the observing frequency $\nu$ into the comoving radial distance $\chi$. We neglect the possible one-halo contribution from extended [\ion{C}{2}] emission around galaxies, which might be challenging to be distinguished from the shot-noise component (but see Section~\ref{sec:discussion} and \citealt{Zhang_2023}).

\begin{figure*}[ht!]
\centering
\includegraphics[width=\textwidth]{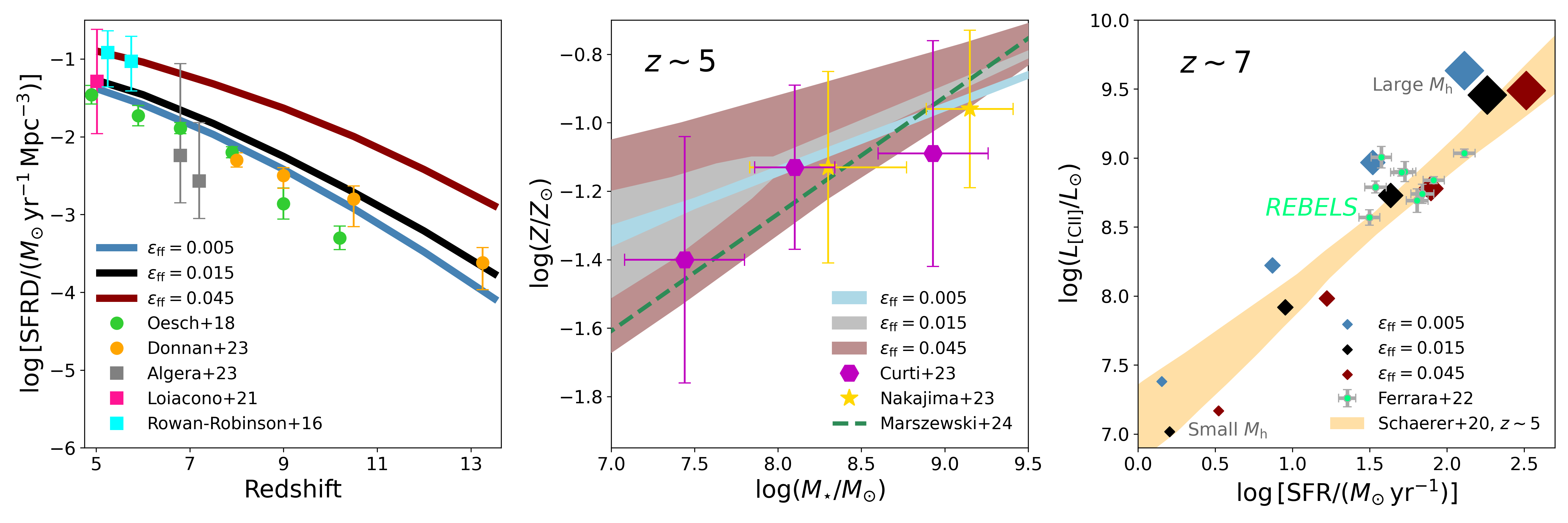}
\caption{Comparison of observables predicted by our analytic models with constraints from recent observations. Left: the cosmic star formation rate density (SFRD) in model variations with different values of the cloud-scale SFE (blue solid: $\epsilon_{\mathrm{ff}} = 0.005$; black solid: $\epsilon_{\mathrm{ff}} = 0.015$; red solid: $\epsilon_{\mathrm{ff}} = 0.045$) and constraints from observations of high-$z$ galaxies at UV \cite[circle:][]{Oesch_2013,Bouwens_2016,Oesch_2018,Donnan_2023} and FIR/mm-wave \cite[square:][]{RR_2016,Loiacono_2021,Algera_2023} wavelengths. Middle: comparison of the stellar mass--metallicity relation (MZR) predicted by the model variations and constraints from recent JWST observations \citep{Curti_2023, Nakajima_2023}. The increased scatter in the MZR at the low-mass end is associated with the increased burstiness of star formation. Also shown is the MZR predicted by high-$z$ FIRE-2 simulations \citep{Marszewski_2024}. Right: comparison between our predicted $L_{\mathrm{[CII]}}$--SFR relation at $z \sim 7$ and recent constraints from ALMA REBELS survey \citep{Ferrara_2022}. Also shown is the best-fit $L_{\mathrm{[CII]}}$--SFR relation for galaxies at $4<z<6$ from the ALMA ALPINE survey \citep{Schaerer_2020}, which suggests little evolution from the relation at $z\sim0$. The four sets of data points in blue, black, and red represent our model predictions for halo masses $\log(M_{\mathrm{h}}/M_{\odot}) \sim 10.5, 11, 11.5$, and 12, respectively (from left to right).}
\label{fig:mzr}
\end{figure*}

\section{Results} \label{sec:result}

\subsection{Properties of Bursty [\ion{C}{2}]-emitting Galaxies} \label{sec:fig1}

Solving the system of DDEs as introduced in Section~\ref{sec:bursty}, we first investigate the evolution of a single halo with an initial mass of $M_{\mathrm{h}} \sim 10^7 M_{\odot}$ at $z = 25$, which evolves into a halo of $M_{\mathrm{h}} \sim 10^{11} M_{\odot}$ by $z = 5$. This is a typical halo that hosts a galaxy contributing to the [\ion{C}{2}] LIM signals at redshifts of interest to us. For comparison, we also solve the same differential equations but without considering the time delay (i.e., $t_\mathcal{D} = 0$) for the same halo.

The left panel of Figure~\ref{fig:check} compares the $L_\mathrm{[CII]}$ evolution of the example halo between the bursty and quasi-equilibrium models and under different assumptions of the cloud-scale SFE, $\epsilon_\mathrm{ff}$. Our fiducial model adopts $\epsilon_{\mathrm{ff}} = 0.015$, consistent with the value determined from the empirical Kennicutt–Schmidt law in local molecular clouds (e.g., \citealt{FM_2022}, and references therein). In light of recent observations that find a large fraction of high-$z$ galaxies with signs of highly bursty star formation, we consider a model variation with $\epsilon_{\mathrm{ff}} = 0.045$, which gives more bursty SFH and may apply to galaxies with higher surface densities in the high-$z$ universe. Furthermore, we also consider the case with $\epsilon_{\mathrm{ff}} = 0.005$ to show how the bursty model reduces to the equilibrium model provided inefficient star formation (and thus weaker feedback). Note that for $\epsilon_{\mathrm{ff}} = 0.045$ we choose to reduce the normalization factor $\eta_{0}$ from 5 to 2.5 in order to (1) avoid overproducing cosmic star formation (see the left panel of Figure~\ref{fig:mzr}) and (2) broadly match high-$z$ UV luminosity functions after accounting for SFR fluctuations, whereas for $\epsilon_{\mathrm{ff}} = 0.005$ we take $\eta_{0} = 5$ as in the fiducial model.

These results demonstrate in general that the higher the $\epsilon_{\mathrm{ff}}$, the more extended in time the transition from bursty to steady star formation (as traced by the [\ion{C}{2}] emission), the stronger the $L_\mathrm{[CII]}$ fluctuations at a given halo mass, and the shorter the duration of each individual starburst cycle. This can be physically interpreted by that more efficient star formation allows the SFR to ``overshoot'' higher before feedback kicks in, but then sources stronger feedback that shuts off star formation (and thus [\ion{C}{2}] emission) more rapidly and requires a higher mass (i.e., a deeper gravitational potential) to stabilize. We choose $\epsilon_\mathrm{ff}$ to be the parameter to vary for model variations given that it strongly affects the overall burstiness of star formation and thus the amplitude of $L_\mathrm{[CII]}$ fluctuations, although other parameters such as $t_\mathcal{D}$ and $\eta$ also play a role (see also \citealt{FM_2022}). The transition to steady star formation stabilizes $L_\mathrm{[CII]}$ at high halo masses and later cosmic times when the rate of mass accretion dominates over the rate of galactic winds driven by supernova feedback. This may also be interpreted as the result of a deep enough gravitational potential that prevents the gas from being evacuated by feedback. For a given halo, the time variation of $L_{\mathrm{[CII]}}$ is roughly symmetric around the equilibrium solution in logarithmic space. This implies that the log-normal PDF defined in Equation~(\ref{eq:pdf}) is indeed a proper choice to describe the $L_{\mathrm{[CII]}}$ distribution.

\begin{figure*}[ht!]\centering
\includegraphics[width=\textwidth]{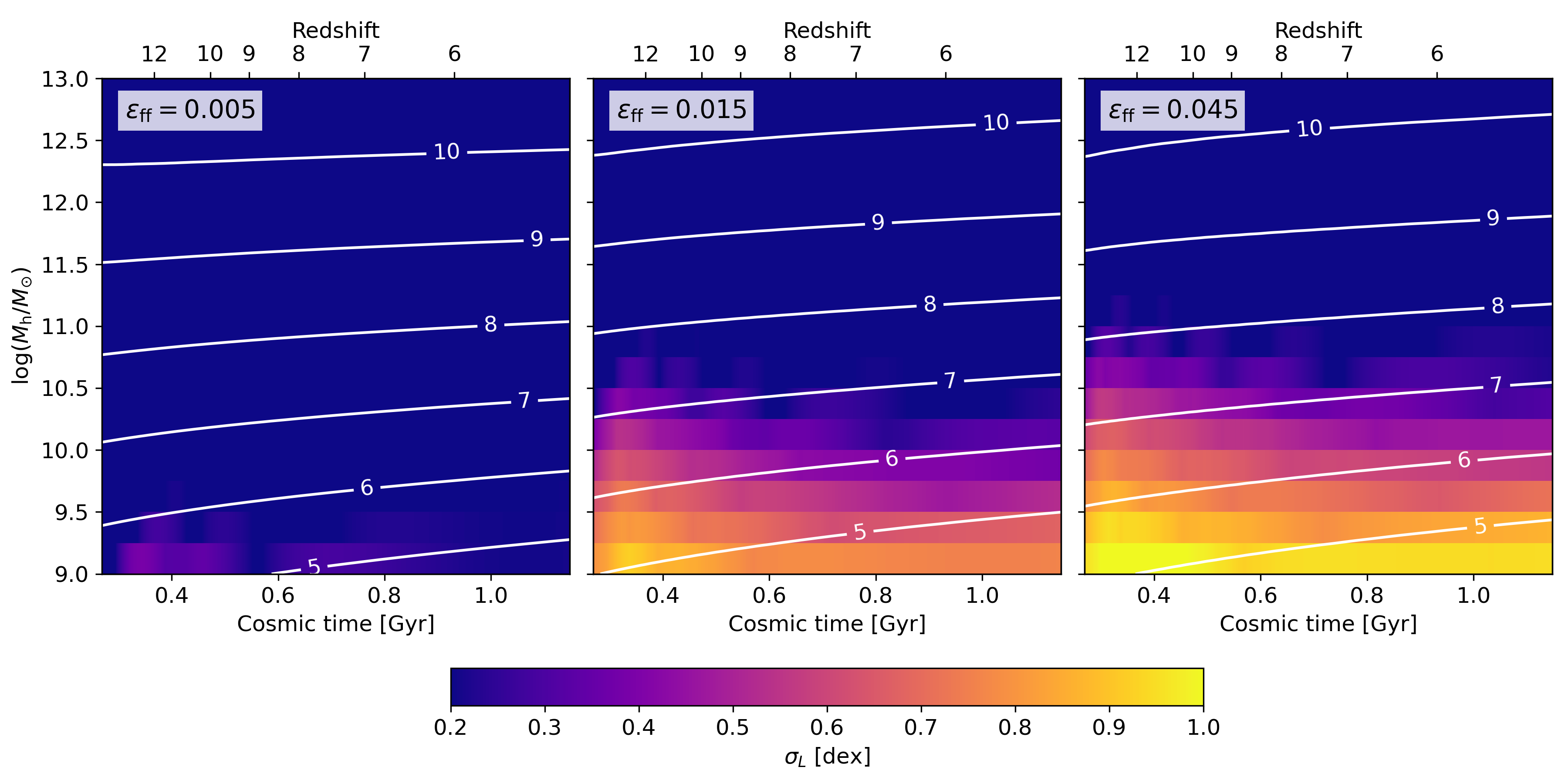}

\caption{The mass and time/redshift dependence of the scatter $\sigma_{L}$ (color code) and the mean \textbf{of logarithmic [\ion{C}{2}] luminosity $\langle \log(L/L_{\odot}) \rangle$} (overlaid contours) obtained by fitting a log-normal distribution to the predicted $L_\mathrm{[CII]}(M_\mathrm{h}, z)$ relation of our bursty star formation model. The 3 panels show results for $\epsilon_{\mathrm{ff}} = 0.005$, 0.015, and 0.045, respectively. The growth of $\sigma_{L}$ toward lower halo masses reflects the increasing burstiness of star formation in halos with shallower gravitational potentials, whereas the tilt of the \textbf{$\langle \log L \rangle$} contours results from the higher growth rate of dark matter halos at earlier times.}
\label{fig:sigma}
\end{figure*}

Detailed properties of the example halo in the fiducial model are shown in the right panel of Figure~\ref{fig:check}, with an emphasis on their highly time-varying evolution compared to the equilibrium scenario. As described by our model, the SFR strictly traces $M_{\mathrm{g}}$ and thus they both follow the same evolution track. Notably, $L_{\mathrm{[CII]}}$ follows the SFR variations in general but in detail has an increased amplitude and a time lag. $L_{\mathrm{[CII]}}$ fluctuates more strongly than the SFR because it scales as the product of $M_{\mathrm{g}}$ and $Z$ and thus reflects their combined effect. The time lag, on the other hand, is associated with the fact that the gas metal enrichment of lags behind star formation --- from Equation~(\ref{eq:metal_ode}) the metal content must peak before $M_{\star}$ but after the SFR. 

By exploring a wide range of initial halo masses and evolving them over time, we can model and characterize the formation histories of the entire high-$z$ galaxy population. Similar to how $L_{\mathrm{[CII]}}$ is treated, we can approximate other quantities like the SFR also with a log-normal distribution. This then allows us to compare our model predictions against various kinds of galaxy observations, including many recent ones enabled by JWST and the Atacama Large Millimeter Array (ALMA).

The left panel of Figure~\ref{fig:mzr} shows a comparison between the star formation rate density (SFRD) in different variations of our model (all integrated down to an SFR of $1\,M_{\odot}/\mathrm{yr}$) and constraints from observations, including data at rest-frame UV \citep{Oesch_2013,Bouwens_2016,Oesch_2018,Donnan_2023} and infrared \citep{RR_2016,Loiacono_2021,Algera_2023} wavelengths. It is clear that a larger value of $\epsilon_{\mathrm{ff}}$ leads to a higher cosmic SFRD because the increased burstiness elevates the (linear) average SFR of galaxies. It has been recently suggested that this effect may help explain the high UV luminosity density observed by JWST at $z \gtrsim 10$ \citep{Shen_2023}. Interestingly, the SFRD from our fiducial model where bursty star formation becomes pronounced for galaxies with $M_\mathrm{h} \lesssim 10^{10}\,M_{\odot}$ (see Figure~\ref{fig:sigma}) provides a good match to the SFRD at $z\gtrsim10$ determined from recent JWST observations (e.g., \citealt{Donnan_2023}). We note though that the SFRD constraints from \citet{Donnan_2023} are based on the photometric redshift and thus may be subject to low-redshift contamination.

\begin{figure*}[ht!]
\centering
\includegraphics[width=1.0\textwidth]{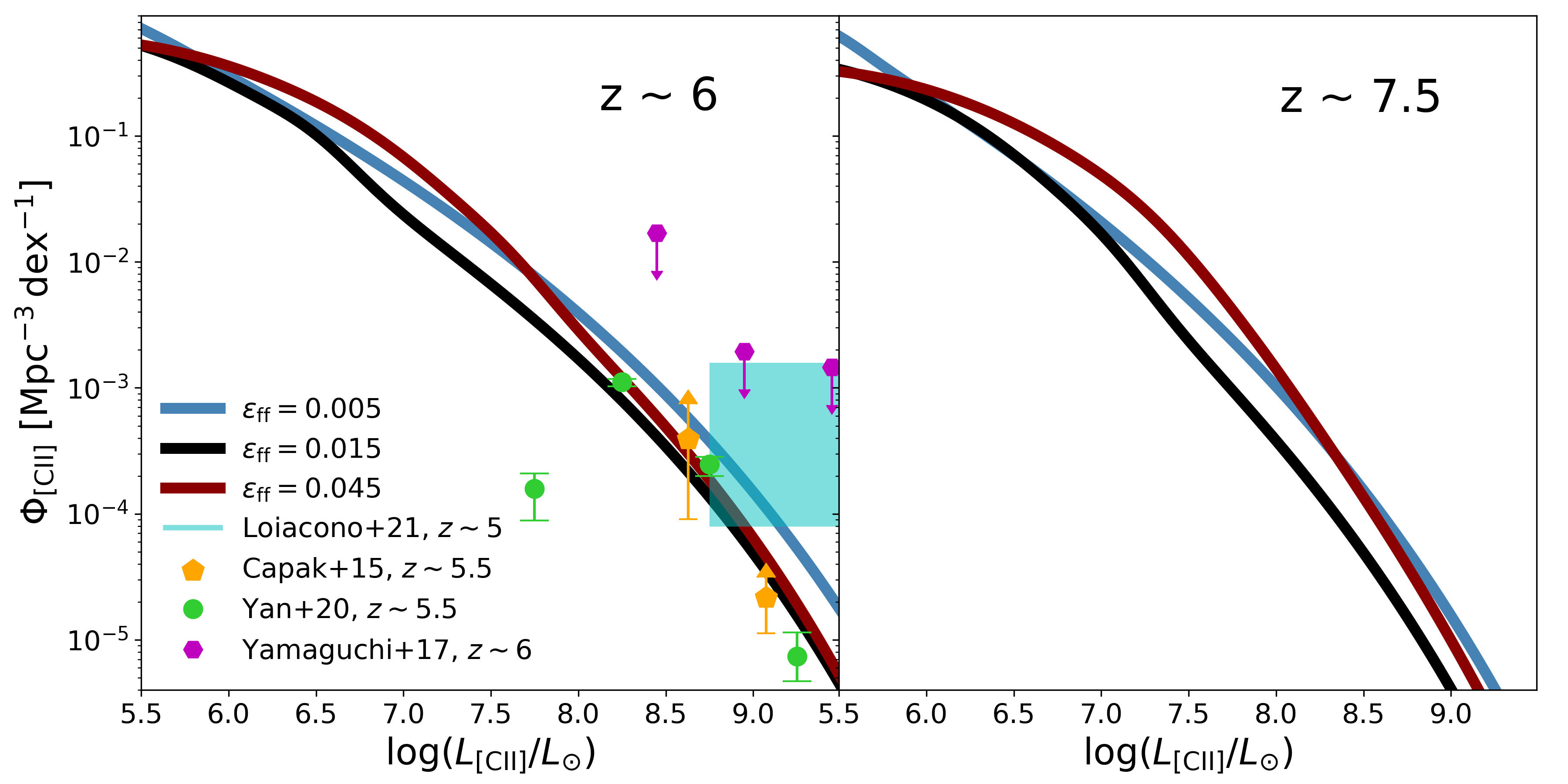}
\caption{The [\ion{C}{2}] luminosity function, $\Phi_\mathrm{[CII]}$, predicted by our models at $z = 6$ (left) and 7.5 (right) for $\epsilon_{\mathrm{ff}} = 0.005$, 0.015, and 0.045. The left panel also shows the comparison with a compilation of constraints from ALMA observations at $z \gtrsim 5$, including both blind surveys \citep{Yamaguchi_2017,Loiacono_2021} and targeted search based on UV-selected samples \citep{Capak_2015,Yan_2020}. The effect of bursty star formation mainly influences the faint end that is difficult to constrain with galaxy observations.}
\label{fig:lf}
\end{figure*}

\begin{figure*}
\centering
\includegraphics[width=1.0\textwidth]{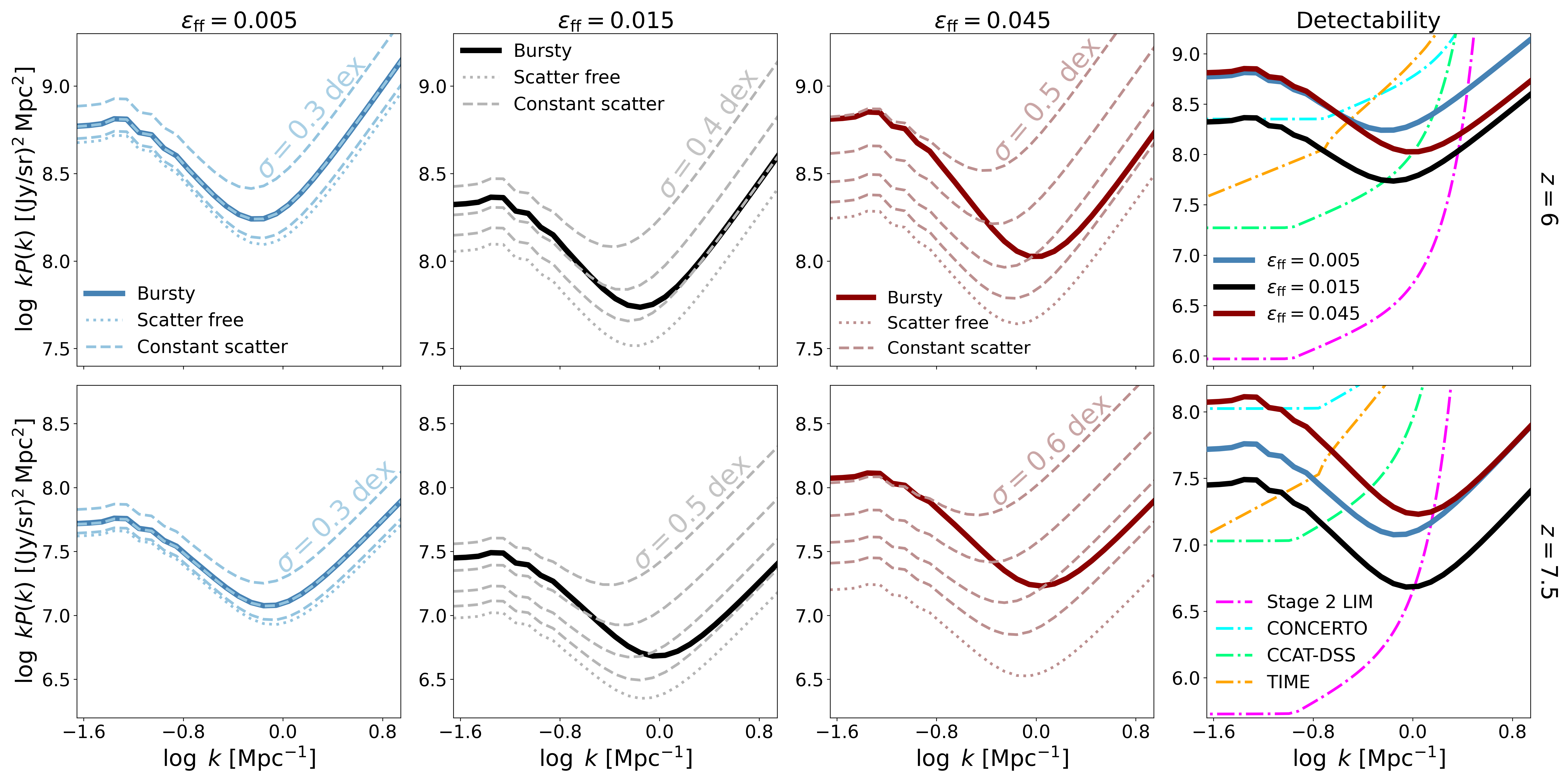}
\caption{The [\ion{C}{2}] power spectra predicted by our analytic framework at $z = 6$ (top) and 7.5 (bottom) for $\epsilon_{\mathrm{ff}} = 0.005$, 0.015, and 0.045. In order to show the scale-dependent distortion of [\ion{C}{2}] power spectra thanks to the burstiness, each panel compares the predictions from our bursty model (solid curve), the scatter-free ($\sigma_{L} = 0$) model (dotted curve), and models assuming a constant $\sigma_{L}$ (dashed curves, from top to bottom each offsets by $-0.1\,$dex from the labeled value). The rightmost column shows the comparison between the predicted [\ion{C}{2}] power spectra with the anticipated sensitivity levels for a number of [\ion{C}{2}] LIM experiments (dash-dotted curves), including CCAT-DSS, CONCERTO, TIME, and a conceptual next-generation experiment (Stage 2 LIM), adapted from \citet{Horlaville_2023}.}
\label{fig:ps}
\end{figure*}

The middle panel of Figure~\ref{fig:mzr} plots the predicted stellar mass--metallicity relations (MZRs) at $z \sim 5$ and a comparison with latest constraints from JWST observations \citep{Curti_2023, Nakajima_2023}. The changing level of burstiness of star formation, induces a scatter in the MZR that increases toward lower masses, as indicated by the different shaded bands. Overall, the MZRs from our models agree well with existing observations (though note the large uncertainties in the measurements derived using mainly the strong-line calibration method), which provide an important sanity check for our modeling of [\ion{C}{2}] luminosity. These results also show that, besides modulating the bursty-to-steady transition of star formation and hence the fluctuations in $Z$ at a given $M_{\star}$, changing $\epsilon_{\mathrm{ff}}$ does not significantly alter the mean MZR. This is because, for individual galaxies, $M_\mathrm{g}$ and $M_{Z}$ are regulated by stellar feedback in similar ways and thus both of them decrease for higher $\epsilon_\mathrm{ff}$, whereas the SFR or $M_{\star}$ is self-regulated by stellar feedback to balance the mass accretion. Thus, in the equilibrium limit, both $Z$ and $M_{\star}$ have little dependence on $\epsilon_\mathrm{ff}$ for a given halo.

The right panel of Figure~\ref{fig:mzr} compares the  $L_{\mathrm{[CII]}}$--SFR relation from our models and the recent ALMA observations \citep{Schaerer_2020, Ferrara_2022}. For clarity, we pick 4 representative halo masses to show the relation predicted by each of our model variations, with the halo mass being represented by the size of the diamond markers. These results indicate that our derived $L_{\mathrm{[CII]}}$--SFR relations are broadly consistent with the existing high-$z$ observations.

\subsection{Predicted [\ion{C}{2}] Summary Statistics} \label{sec:lfps}

An essential step for estimating summary statistics of [\ion{C}{2}] LIM signals is to define the $L_\mathrm{[CII]}$--$M_\mathrm{h}$ relation. As shown in Figure~\ref{fig:check}, a mass- and time-dependent scatter in $L_\mathrm{[CII]}$ naturally arises from bursty star formation. We track the evolution of $L_{\mathrm{[CII]}}$ calculated from halo properties using Equation~(\ref{eq:LCII}) for individual halos over a wide range of $M_{\mathrm{h}}$ to ensure that halos responsible for cosmic star formation and [\ion{C}{2}] emission are well sampled in the redshift range of interest ($5 \lesssim z \lesssim 15$). We then fit a log-normal distribution to capture the range of $L_{\mathrm{[CII]}}$ fluctuations. As explained in Section~\ref{sec:lcii_distr}, when fitting $L_{\mathrm{[CII]}}$ fluctuations, the scatter $\sigma_{L}$ is determined by approximating the envelope of $L_{\mathrm{[CII]}}$ fluctuations derived analytically with the $\pm 1\sigma_{L}$ range, whereas the mean is given by the equilibrium model solution.

Figure~\ref{fig:sigma} shows the mean [\ion{C}{2}] luminosity and its logarithmic scatter for our 3 model variations distinguished by different values of $\epsilon_{\mathrm{ff}}$. The higher [\ion{C}{2}] luminosity fluctuations and the later transition from bursty to steady star formation for a larger $\epsilon_{\mathrm{ff}}$ can be clearly seen from the color coding, whereas the contours show the much weaker dependence of the mean [\ion{C}{2}] luminosity $\langle \log L \rangle$ on $\epsilon_{\mathrm{ff}}$, driven primarily by the gas content. From these 3 panels, it is clear that low-mass, [\ion{C}{2}]-faint galaxies are most affected by bursty star formation --- in the case of $\epsilon_{\mathrm{ff}} = 0.015$ and 0.045, $L_{\mathrm{[CII]}}$ of the lowest-mass halos can be scattered by more than an order of magnitude. Since low-mass galaxies contribute significantly to the aggregate [\ion{C}{2}] emission, this motivates us to search for the imprints of bursty star formation on [\ion{C}{2}] LIM signals.

The predicted [\ion{C}{2}] luminosity functions at $z = 6$ and 7.5 are shown in Figure~\ref{fig:lf}. The left panel shows the comparison with recent observational constraints from ALMA, including blind surveys \citep{Yamaguchi_2017, Loiacono_2021} and targeted searches based on UV-selected samples as lower limits \citep{Capak_2015, Yan_2020}. The right panel only shows our model predictions due to the lack of observations. The impact of [\ion{C}{2}] luminosity fluctuations from bursty star formation on $\Phi_\mathrm{[CII]}$ is clearly visible, especially at the faint end. A larger $\epsilon_{\mathrm{ff}}$ results in a scatter that evolves more strongly with mass, which in turn yields a more elevated faint end with the elevation extending to higher luminosities. Such a net distortion of $\Phi_\mathrm{[CII]}$ is also manifested in the power spectrum (see Figure~\ref{fig:ps}). Note that the lack of constraints on the faint end of $\Phi_\mathrm{[CII]}$ from direct galaxy detections again demonstrates the compelling need for LIM observations.

Using the extracted $L_\mathrm{[CII]}(M_\mathrm{h},z)$ relation, we show in Figure~\ref{fig:ps} the [\ion{C}{2}] power spectra predicted by our models and then compare the prediction of [\ion{C}{2}] power spectra to the expected sensitivity levels of several [\ion{C}{2}] LIM experiments, including CCAT-DSS \citep{Stacey_2018, CCAT_2023}, CONCERTO \citep{CONCERTO_2020, Monfardini_2022}, TIME \citep{Crites_2014, Sun_2021}, and a hypothetical Stage~2 experiment \cite[see][]{Horlaville_2023}. The left 3 columns show [\ion{C}{2}] power spectra derived from our full bursty model (solid curves), scatter-free ($\sigma_{L} = 0$) model (dotted curves), and a few reference models assuming a range of mass-independent scatters (dashed curves). The rightmost column shows the comparison against power spectrum sensitivity levels of the aforementioned [\ion{C}{2}] LIM experiments (dash-dotted curves).

Several key results are apparent from the comparison of these curves:
\begin{itemize}
\item By comparing the solid (mass-dependent scatter) and dotted curves (scatter free), we show that stochastic, scattered $L_{[\mathrm{CII}]}$ induced by bursty star formation can boost the overall [\ion{C}{2}] power spectrum signal level. This, in conjunction with the left panel of Figure~\ref{fig:mzr}, demonstrates that burstiness affects the [\ion{C}{2}] power spectrum on all scales by increasing the cosmic SFRD and thus the [\ion{C}{2}] luminosity density. This is essentially due to the fact that the linear average of a log-normal distribution is larger for a larger scatter. We choose not to fix the linear average by normalization (as done in some other models in the literature), since the functional form used for fitting [\ion{C}{2}] fluctuations is physically grounded in each of our model variations. We also note that the solid curves never fall below the model that assumes a constant 0.2\,dex scatter, given the minimum scatter imposed.

\item By contrasting the solid curve against the dashed curves, we find a pronounced deviation of both the power spectrum amplitude and shape in our bursty model from models assuming a fixed, mass-independent scatter which is often implemented in other models in the literature. The large-scale, clustering power tends to favor a larger scatter that can be more than a factor of 2 higher than the value favored by the small-scale, shot-noise power, while the small-scale, shot-noise power tends to approach the model that assumes a 0.2\,dex scatter. This is naturally explained by (1) $P_{\mathrm{clust}}$ is more sensitive to lower mass halos while $P_{\mathrm{shot}}$ is more sensitive to higher mass halos, (2) the star formation is more bursty and thus leads to larger values of $\sigma_{L}$ in lower mass halos, and (3) high-mass halos host galaxies with more steady star formation and $\sigma_{L}$ closer to the minimum value imposed (see Figure~\ref{fig:sigma}). The power spectrum in the bursty model is thus \textit{incompatible} with that predicted by adding a constant scatter to the intrinsic $L_\mathrm{[CII]}$--$M_\mathrm{h}$ relation. 

\item From the comparison between differences in the power spectrum caused by bursty star formation and the sensitivity levels of various [\ion{C}{2}] LIM experiments, it is evident that the large-scale, clustering power is likely detectable by the first-generation experiments (CCAT-DSS, CONCERTO, and TIME), whereas the small-scale, shot-noise power likely requires better high-$k$ sensitivity that CCAT-DSS starts to offer (e.g., at $z=6$). Stage 2 experiments will ultimately provide the sensitivity to the shape distortion of power spectrum necessary for accurately constraining the effect of bursty star formation.

\item The scale-dependent modulation of power spectrum shape is a non-trivial effect that should be taken into account when modeling and analyzing the high-$z$ [\ion{C}{2}] LIM signals. This emphasizes the importance of considering a mass-dependent scatter associated with bursty star formation at high redshift and the risk of introducing a strong systematic bias if not.

\end{itemize}

\section{Discussion} \label{sec:discussion}

Overall, our simple analytic framework of the bursty star formation and [\ion{C}{2}] LIM predicts shape distortions in the [\ion{C}{2}] (LIM) summary statistics. Varying the level of burstiness can lead to different and potentially distinguishable imprints in the [\ion{C}{2}] power spectrum. We note that model parameters other than $\epsilon_{\mathrm{ff}}$, such as $\eta$ and $t_{\mathcal{D}}$, can also vary the level of the burstiness. However, they do not affect bursty star formation in the same way, which makes them not fully degenerate with each other. Specifically, $t_{\mathcal{D}}$, when varied within a physically plausible range, only affects the burstiness modestly and does not impact the equilibrium, steady-state SFR. $\eta$, on the other hand, can strongly affect the equilibrium, steady-state SFR. Thus, these parameters can be separately constrained by other observables such as the galaxy UV luminosity function \cite[see][for more thorough investigation]{FM_2022}. We also emphasize that while more than one parameters in our model impact the burstiness, our goal in this work is to establish the connection between bursty star formation and the summary statistics of [\ion{C}{2}] LIM rather than parameter inference, which may be most effectively done by combining multiple probes.

It should be also noted that the effects of bursty star formation on LIM observables explored in this work are by no means unique to high-$z$ [\ion{C}{2}] emission --- other common target lines for LIM that trace star-forming galaxies such as H$\alpha$, CO, and [\ion{O}{3}] are also subject to these effects and bursty SFHs widely exist for low-mass galaxies even in the nearby universe.

Several limitations of the current model and analysis are noteworthy, which point to ways to extend the existing framework in future studies. First, our model for bursty star formation is ultimately simple and thus requires further examinations and improvements. Taking into account of additional channels of stellar feedback, such as radiation pressure that can regulate the cloud-scale SFE and the recycling of outflows, might provide more robust descriptions of bursty star formation in high-$z$ galaxies. Fluctuations of the halo accretion rate due to e.g., mergers, on the other hand, can also induce burstiness, but this may be a minor effect since the mass accretion rate typically has a modest $\sim 0.3$\,dex scatter and mergers do not contribute significantly to the growth for early galaxies \cite[see][and references therein]{FM_2022}. It is also possible for the significantly denser and less metal-enriched high-$z$ gas clouds to have starbursts free of any feedback, which give rise to a different regime of bursty SFHs than that considered in this work \cite[see][]{Dekel_2023,Li_2023}. We defer to future work a more thorough analysis of different mechanisms of bursty star formation.

Second, many factors that could influence the [\ion{C}{2}] emission are not inspected in this work. For simplicity, we ignore variations of $f_\mathrm{[CII]}$ and its possible connection to bursty star formation, though we note that the $f_\mathrm{[CII]}$ could be larger in low-metallicity systems (see \citealt{Madden_2020} for observations on local low-metallicity dwarfs) due to the fact that the photoionized carbon can penetrate deeper in the low-metallicity, less-dusty clouds and create thicker [\ion{C}{2}] layers around the clouds, thus leading to stronger [\ion{C}{2}] emission. We also ignore the effect of cosmic microwave background (CMB) on [\ion{C}{2}] emission as both an additional heating source and a background radiation against which [\ion{C}{2}] is observed \citep{Lagache_2018, Liang_2023}. More accurate [\ion{C}{2}] emission models that account for these effects as well as the changing interstellar and cosmic ray radiation fields, dust content, etc. are worthwhile to develop in future studies \citep{Ferrara_2019, Liang_2023}. 

Moreover, we note that the effects of bursty star formation on summary statistics of the [\ion{C}{2}] emission explored here might be degenerate with several astrophysical factors, including a different intrinsic scaling relation of $L_{[\mathrm{CII}]}$ and $M_\mathrm{h}$ (i.e., the light-to-mass ratio) and the [\ion{C}{2}] deficit commonly observed in IR-luminous galaxies \citep{Liang_2023}. Disentangling effects of a mass-dependent scatter from all these factors to uniquely constrain bursty star formation may be challenging using only LIM observations. Additional information about the stochasticity of the light-to-mass ratio may be obtained from the clustering of bright emitters \cite[e.g.,][]{Munoz_2023}. 

Lastly, when modeling the [\ion{C}{2}] power spectrum we do not include the one-halo term, which in principle can complicate the power spectrum shape analysis at intermediate scales. The [\ion{C}{2}] halo emission discovered at high redshift that extends out to the circumgalactic medium scales ($\sim10\,$kpc) may source a non-negligible one-halo component of the [\ion{C}{2}] LIM power spectrum \citep{Fujimoto_2019}, though we note that \cite{Zhang_2023} find that the one-halo term might be overwhelmed by the shot-noise term. We plan to model this extended [\ion{C}{2}] emission self-consistently with recipes of feedback-regulated bursty star formation and galactic inflows/outflows in future work.

\section{Conclusions} \label{sec:summary}

For the first time, we investigate how bursty star formation driven by delayed supernova feedback may impact the observed [\ion{C}{2}] LIM signals of galaxies at EoR. We approach this problem with an analytic framework that bridges high-$z$ galaxy formation with bursty SFHs and the production of [\ion{C}{2}] LIM signals. We show that bursty star formation provides a natural way to introduce stochasticity into the $L_{\mathrm{[CII]}}$--$M_\mathrm{h}$ relation by yielding a strongly mass-dependent scatter. We have experimented with various values of the cloud-scale SFE, $\epsilon_{\mathrm{ff}}$, as the knob for adjusting the level of burstiness in order to characterize the impact of bursty star formation on summary statistics of [\ion{C}{2}] LIM signals, namely the [\ion{C}{2}] luminosity function and power spectrum. 

We find that our predicted [\ion{C}{2}] luminosity functions are in good agreement with recent constraints from ALMA. With our bursty star formation model, we demonstrate that the transition from bursty to steady star formation can significantly distort the faint-end [\ion{C}{2}] luminosity function and the shape of [\ion{C}{2}] power spectrum. Some of these imprints, such as the elevated faint end of $\Phi_\mathrm{[CII]}$ and the boosted $P_{\mathrm{clust}}$, are at the level of being detectable even by the first-generation high-$z$ [\ion{C}{2}] LIM experiments, and thus non-negligible for both modeling and analysis. Our results shed light on the usefulness of LIM to statistically constrain bursty star formation, as a complementary method to detailed observations of individual galaxies. We emphasize that the majority of existing observations can only probe the most luminous [\ion{C}{2}]-emitting galaxies at the redshifts considered in this work, whereas the fainter and more bursty population can be efficiently probed by [\ion{C}{2}] LIM experiments.

In summary, bursty star formation deserves particular attention in the modeling and analysis of LIM signals of high-$z$ star-forming galaxies given the strongly mass-dependent scatter in line luminosity it can induce. Using [\ion{C}{2}] as an example, results from this work pave the way for exploring the astrophysical significance and implications of this phenomenon. Characteristic imprints of bursty SFHs on [\ion{C}{2}] LIM summary statistics, particularly the scale-dependent boost to the [\ion{C}{2}] power spectrum amplitude, provide a unique quantification of bursty star formation and the physics behind it, which will be leveraged by forthcoming experiments to further our understanding of high-$z$ galaxy formation. 

\section*{Acknowledgement}
We thank the referee, Matthieu B\'{e}thermin, for valuable comments that helped improve this paper. We are indebted to Dongwoo Chung and Andrea Ferrara for helpful comments on an early draft. We also thank Dongwoo Chung and Patrick Horlaville for sharing the sensitivity estimates of [\ion{C}{2}] LIM surveys, as well as Claude-Andr\'{e} Faucher-Gigu\`{e}re, Andrew Hearin, Adam Lidz, and Norm Murray for stimulating discussions. G.S. was supported by a CIERA Postdoctoral Fellowship. T.-C.C. acknowledges support by NASA ROSES grant 21-ADAP21-0122. Part of this work was done at Jet Propulsion Laboratory, California Institute of Technology, under a contract with the National Aeronautics and Space Administration. S.R.F. was supported by NASA through award 80NSSC22K0818 and by the National Science Foundation through award AST-2205900. This work was initiated in part at Aspen Center for Physics, which is supported by National Science Foundation grant PHY-2210452. 

\software{astropy \citep{2013A&A...558A..33A,2018AJ....156..123A,2022ApJ...935..167A},   
          hmf \citep{2013A&C.....3...23M}, 
          matplotlib \citep{Hunter:2007}, 
          numpy \citep{harris2020array}, 
          scipy \citep{2020SciPy-NMeth}
          }

\begin{figure*}
\centering
\includegraphics[width=\textwidth]{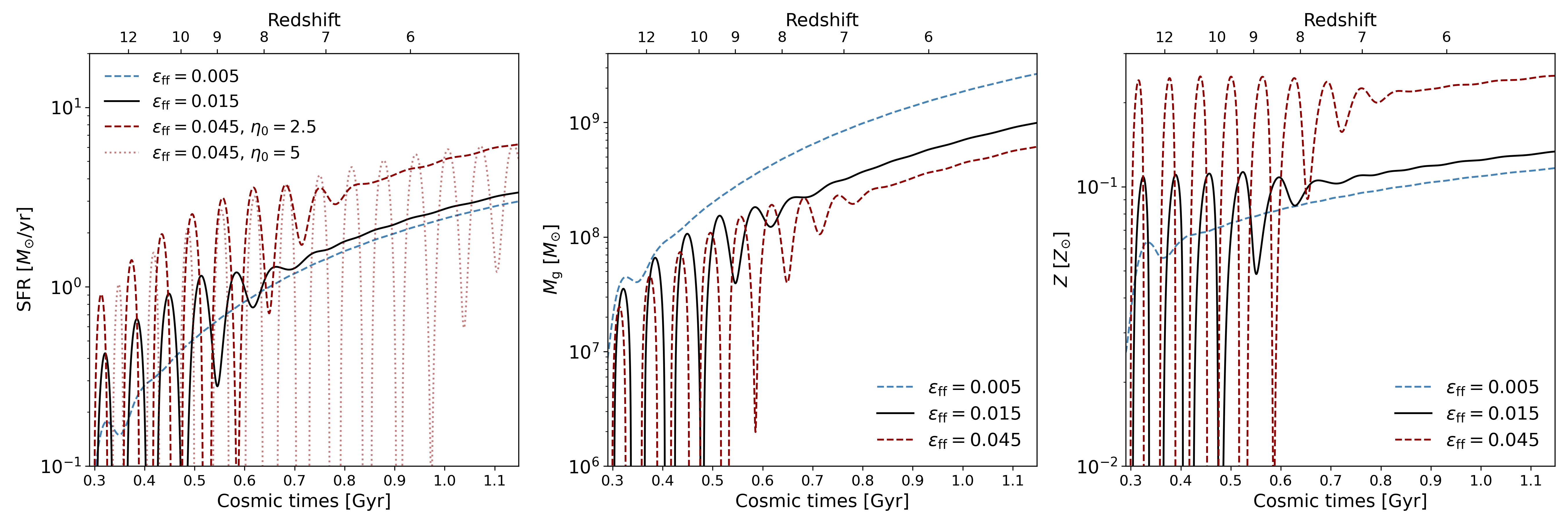}
\caption{Comparison of the galaxy properties for the same example halo shown in Figure~\ref{fig:check}.
Left: star formation histories. To show the reason for adopting a different $\eta_{0} = 2.5$ for the case with $\epsilon_{\mathrm{ff}} = 0.045$, we also plot the case where $\eta_{0} = 5$ is assumed (as in the two other models) which predicts excessive burstiness. Middle: evolution of the gas mass $M_{\mathrm{g}}$. Right: evolution of the gas metallicity $Z$.}
\label{fig:vary}
\end{figure*}

\begin{figure*}
\centering
\includegraphics[width=\textwidth]{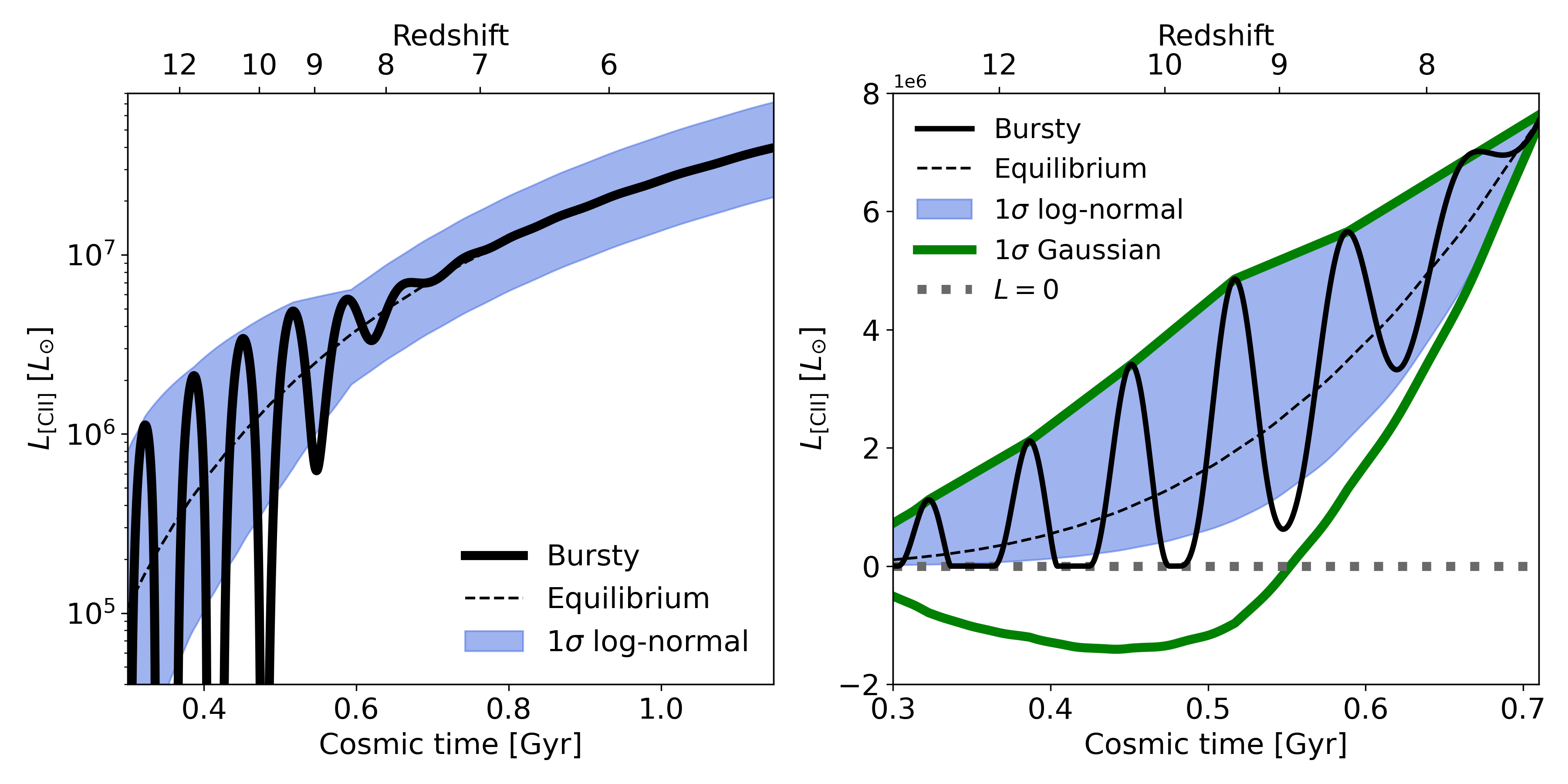}
\caption{Left: time evolution of [\ion{C}{2}] luminosity of the same example halo shown in Figure~\ref{fig:check} fit by a log-normal distribution. The blue shaded band indicates the $\pm 1\sigma$ range. The fiducial value $\epsilon_{\mathrm{ff}} = 0.015$ is assumed. Right: comparison of fitting [\ion{C}{2}] luminosity fluctuations with a log-normal distribution vs. a Gaussian distribution.}
\label{fig:lognormal}
\end{figure*}

\appendix

\section{Evolution of galaxy properties in the bursty star formation model}

For completeness, we show in Figure~\ref{fig:vary} the evolution of several galaxy properties of the same example halo considered in Figure~\ref{fig:check}. The left panel shows star formation histories predicted by our model variations. The overall trend is similar to that of the [\ion{C}{2}] luminosity shown in the left panel of Figure~\ref{fig:check}, where the higher the cloud-scale SFE ($\epsilon_{\mathrm{ff}}$), the more extended in time the transition from bursty to steady star formation, the stronger the SFR fluctuations at a given halo mass, and the shorter the duration of each individual burst cycle. Note that the steady-state SFR (i.e., the equilibrium solution) is barely sensitive to $\epsilon_\mathrm{ff}$ and mainly set by $\eta$, with higher $\eta$ corresponding to stronger stellar feedback and thus lower SFR. Here, we also include a comparison of models with $\epsilon_{\mathrm{ff}} = 0.045$ and different values of $\eta_0$. It is clear that assuming $\epsilon_{\mathrm{ff}} = 0.045$ and $\eta_0 = 5$ (as assumed in the two other models) leads to excessive burstiness that is found to significantly overproduce the SFRD. Setting $\eta_0$ to 2.5 instead predicts less amount of burstiness and the resulting SFRD is in better agreement with observations despite the overall higher steady-state SFR. The middle and right panels show the evolution of gas mass ($M_\mathrm{g}$) and metallicity ($Z$), respectively. We emphasize that in our model the physics of [\ion{C}{2}] production is most directly related to $M_\mathrm{g}$ and $Z$ rather than the SFR. Changing $\epsilon_{\mathrm{ff}}$ affects the steady-state $M_\mathrm{g}$ and $Z$ in opposite ways (see the middle and right panel of Figure~\ref{fig:vary}), though the impact on the former is more substantial. This explains the overall higher [\ion{C}{2}] luminosity in the $\epsilon_{\mathrm{ff}}=0.005$ model as shown in the left panel of Figure~\ref{fig:check}.

\section{Approximating [\ion{C}{2}] luminosity fluctuations with a log-normal distribution}

We show in Figure~\ref{fig:lognormal}, which considers the same example halo as investigated in Figure~\ref{fig:check} ($\epsilon_{\mathrm{ff}} = 0.015$), that one can use a log-normal distribution to approximate the [\ion{C}{2}] luminosity fluctuations derived from our analytic bursty star formation model for the purpose of predicting [\ion{C}{2}] LIM signals. Because the fluctuations asymptote to the equilibrium $L_{\mathrm{[CII]}}$ solution, the mean of the distribution is set by physics rather than chosen by hand, whereas the $1\sigma$ width of the distribution is determined by fitting the peaks of individual bursts. As shown by the envelope in the left panel, the log-normal fit captures the roughly symmetric fluctuations around the equilibrium solution in logarithmic space. We have also verified that the slightly different choices of other parameters, such as $\eta_0$, $t_{\mathcal{D}}$, and the initial time of halo evolution ($z_{\mathrm{initial}}$), result in $L_{\mathrm{[CII]}}$ evolution tracks that largely fall within the $\pm1\sigma$ range of the envelope. While the best-fit $1\sigma$ envelope may not be able to capture extreme $L_{\mathrm{[CII]}}$ values that approach 0, we note that these deviations are insignificant for the [\ion{C}{2}] LIM signals of interest. For comparison, in the right panel we show how the log-normal distribution outperforms a Gaussian distribution determined in a similar way. In the case of small mean $L_{\mathrm{[CII]}}$ and large scatter, the Gaussian distribution pushes a significant fraction of the probability distribution to the unphysical $L_{\mathrm{[CII]}} < 0$ regime.

\bibliography{sample631}{}
\bibliographystyle{aasjournal}

\end{document}